\documentclass[graybox]{svmult}

\usepackage{amsmath,mathptmx}       
\usepackage{helvet}         
\usepackage{courier}        
\usepackage{type1cm}        
\usepackage{makeidx}         
\usepackage{graphicx}        
\usepackage{multicol}        
\usepackage[bottom]{footmisc}
\usepackage{amsfonts}	
\usepackage{hyperref,cite}

\makeindex             

\newcommand{\beqs}{\begin{equation*}}

\newcommand{\beq}{\begin{equation}}

\newcommand{\eeqs}{\end{equation*}}

\newcommand{\eeq}{\end{equation}}

\newcommand{\beqas}{\begin{eqnarray*}}

\newcommand{\beqa}{\begin{eqnarray}}

\newcommand{\eeqas}{\end{eqnarray*}}

\newcommand{\eeqa}{\end{eqnarray}}

\newcommand{\nn}{\nonumber}

\newcommand{\eq}[2]{\begin{equation} #1 \label{#2} \end{equation}}

\newcommand{\al}{\alpha}

\newcommand{\be}{\beta}

\newcommand{\ga}{\gamma}

\newcommand{\de}{\delta}

\newcommand{\om}{\omega}

\newcommand{\si}{\sigma}

\newcommand{\Ga}{\Gamma}

\newcommand{\blist}{\begin{itemize}}

\newcommand{\elist}{\end{itemize}}

\providecommand{\href}[2]{#2}

\DeclareFontFamily{OT1}{rsfs}{}

\DeclareFontShape{OT1}{rsfs}{m}{n}{ <-7> rsfs5 <7-10> rsfs7 <10->rsfs10}{} 

\DeclareMathAlphabet{\mycal}{OT1}{rsfs}{m}{n}

\DeclareMathOperator{\extdm}{d}

\newcommand{\extd}{\extdm \!}

\newcommand{\tr}{\textrm{tr}}

\newcommand{\re}[1]{(\ref{#1})}

\begin{document}

\title*{Holographic Chern--Simons Theories}
\authorrunning{D.~Grumiller et al.}
\author{H.~Afshar, A.~Bagchi, S.~Detournay, \underline{D.~Grumiller}, S.~Prohazka and M.~Riegler}
\institute{Hamid Afshar \at Centre for Theoretical Physics, University of Groningen, Nijenborgh 4, 9747 AG Groningen, The
Netherlands, \email{h.r.afshar@rug.nl}
\and Arjun Bagchi \at Indian Institute of Science Education and Research, Pune 411008, India,
\email{a.bagchi@iiserpune.ac.in}
\and St\'ephane Detournay \at Physique Th\'eorique et Math\'ematique,
Universit\'e Libre de Bruxelles and International Solvay Institutes, Campus Plaine C.P.~231, B-1050 Bruxelles, Belgium, \email{sdetourn@ulb.ac.be}
\and Daniel Grumiller \at Institute for Theoretical Physics, Vienna University of Technology, Wiedner Hauptstrasse 8-10/136, A-1040 Vienna, Austria, \email{grumil@hep.itp.tuwien.ac.at}
\and Stefan Prohazka \at Institute for Theoretical Physics, Vienna University of Technology, Wiedner Hauptstrasse 8-10/136, A-1040 Vienna, Austria, \email{prohazka@hep.itp.tuwien.ac.at}
\and Max Riegler \at Institute for Theoretical Physics, Vienna University of Technology, Wiedner Hauptstrasse 8-10/136, A-1040 Vienna, Austria, \email{rieglerm@hep.itp.tuwien.ac.at} 
}
%
%
\maketitle


\abstract{%
Chern--Simons theories in three dimensions are topological field theories that may have a holographic interpretation for suitable chosen gauge groups and boundary conditions on the fields. Conformal Chern--Simons gravity is a topological model of 3-dimensional gravity that exhibits Weyl invariance and allows various holographic descriptions, including Anti-de~Sitter, Lobachevsky and flat space holography. The same model also allows to address some aspects that arise in higher spin gravity in a considerably simplified setup, since both types of models have gauge symmetries other than diffeomorphisms. In these lectures we summarize briefly recent results.  
}

\section{Introduction}
\label{sec:1}

Chern--Simons theories in three dimensions have a wide range of applications in mathematics and physics (see \cite{Dunne:1998qy,Marino:2004uf,Zanelli:2005sa,Zanelli:2008sn,Marino:2011nm,Zanelli:2012px,Bergshoeff:2014bia} for various reviews).
The bulk action
\eq{
S_{\textrm{\tiny CS}}[A]=\frac{k_{\textrm{\tiny CS}}}{4\pi}\,\int_{\cal M}\,\tr \big(A\wedge \extd A + \tfrac23\,A\wedge A\wedge A\big)
}{eq:paros1}
depends on a dimensionless coupling constant, the Chern--Simons level $k_{\textrm{\tiny CS}}$, a Lie-algebra valued connection 1-form $A$ and a manifold $\cal M$ that often has some boundary $\partial\cal M$. In these lectures we always assume that ${\cal M}$ topologically is either a filled cylinder or a filled torus.

While the Lagrange-3-form in the action \eqref{eq:paros1} is not gauge invariant, the equations of motion are gauge invariant,
\eq{
F = \extd A + A\wedge A = 0\,,
}{eq:paros2}
and show that locally all solutions are pure gauge. The theory is topological in the sense that its action does not depend on the metric, and also topological in the sense that the theory has no local physical degrees of freedom (see \cite{Birmingham:1991ty} for a review on topological field theories).

Thus, all physical excitations are of global nature, and if ${\cal M}$ has a boundary one can picture the excitations as edge states localized at the boundary, much like in the Anti-de~Sitter/conformal field theory (AdS/CFT) correspondence.

The precise boundary conditions imposed on the connection $A$ are a crucial input in the specification of the model, and the same bulk action can describe completely different physical systems, depending on the specific choice of boundary data.

Prominent examples of Chern--Simons theories with special boundary conditions are Einstein gravity with negative cosmological constant \cite{Achucarro:1987vz,Witten:1988hc} and higher spin theories \cite{Henneaux:2010xg,Campoleoni:2010zq}, some aspects of which are reviewed below.

In these lectures we focus mostly on a specific theory of gravity, conformal Chern--Simons gravity (CSG) \cite{Deser:1982vy,Deser:1982wh,Deser:1982a}. Its bulk action is similar to the Chern--Simons action \eqref{eq:paros1}, but depends on a connection that is not a fundamental field, namely on the Christoffel connection.
\eq{
S_{\textrm{\tiny CSG}} = \frac{k}{4\pi}\,\int_{\cal M} \extd^3x\,\epsilon^{\al\be\ga}\,\Ga^\mu{}_{\al\nu}\,\big(\partial_\be \Ga^\nu{}_{\ga\mu}+\tfrac23\,\Ga^\nu{}_{\be\si}\Ga^\si{}_{\ga\mu}\big)
}{eq:paros3}
Consequently, the equations of motion obtained by varying the action \eqref{eq:paros1} with respect to the metric do not imply flatness of the geometry, but only conformal flatness.
\eq{
C_{\mu\nu} = \tfrac12\,\varepsilon_\mu{}^{\al\be}\nabla_\al R_{\be\nu} + (\mu\leftrightarrow\nu) = 0
}{eq:paros4}
The quantity $C_{\mu\nu}$ is the Cotton tensor, which vanishes in three dimensions if and only if spacetime is conformally flat (see for instance \cite{Garcia:2003bw}).

Thus, as opposed to 3-dimensional Einstein gravity with negative cosmological constant, which allows only locally AdS solutions and thus only AdS holography, CSG has also some non-AdS solutions and is thus a simple model that allows to study non-AdS holography. Moreover, CSG has an additional gauge symmetry, namely Weyl symmetry 
\eq{
g_{\mu\nu}\to e^{2\Omega}\,g_{\mu\nu}\,, 
}{eq:angelinajolie}
so that metrics that are not diffeomorphic to each other can nevertheless be gauge equivalent. All these properties are shared by higher spin gravity, which is why CSG can be regarded as a simple toy model for higher spin gravity and non-AdS holography (see \cite{Gary:2012ms,Afshar:2012nk} for the higher spin perspective and \cite{Afshar:2011yh,Afshar:2011qw} for the CSG perspective).

We address now which boundary conditions are possible in CSG. In principle, any conformally flat metric is an allowed background. However, for practical applications it usually makes sense to consider backgrounds that have at least one Killing vector, e.g., associated with asymptotic time translations. In that case, a Kaluza--Klein reduction to two dimensions reduces CSG to a specific non-linear Maxwell--Einstein theory \cite{Guralnik:2003we}. This theory in turn can be mapped to a specific Dilaton--Maxwell--Einstein theory, whose classical solutions can be
found globally \cite{Grumiller:2003ad}. It turns out that all such solutions have additional Killing vectors: they are either maximally symmetric, i.e., have six Killing vectors, or they have four Killing vectors.

The first option allows to study AdS holography, flat space holography and de~Sitter holography. The second option allows to study Lobachevsky holography. In the rest of these lectures we review some of these holographic setups and recent results. In section \ref{sec:2} we review AdS holography. In section \ref{sec:3} we address Lobachevsky holography. In section \ref{sec:4} we focus on flat space holography, in particular in the context of quantum gravity toy models.

\section{Anti-de~Sitter holography}
\label{sec:2}


Holography provides a map between quantum gravity in $d+1$ dimensions and quantum field theories in $d$ dimensions. While holographic correspondences exist that involve specific types of non-unitary theories --- see \cite{Grumiller:2008qz,Grumiller:2013at} and references therein --- for many purposes one would like to insist on unitarity.

As we shall review in sections \ref{sec:2.1} and \ref{sec:4.2}, in 3-dimensional gravity unitarity prefers spacetimes with AdS asymptotics for quantization of parity even theories and asymptotically flat spacetimes for quantization of parity odd theories.
There are two pure gravity models without local degrees of freedom in three dimensions, parity even
{\it Einstein-Hilbert gravity} (EHG) and parity odd {\it conformal Chern-Simons gravity} (CSG). 
These models can be written as Chern-Simons topological gauge theories of level $k_{\textrm{\tiny CS}}$ for SO(2,2) AdS \cite{Achucarro:1986vz,Witten:1988hc} and SO(3,2) conformal \cite{Horne:1988jf} groups respectively, with a {\it proper }non-degenerate bilinear form. 
The AdS algebra 
\begin{align}\label{algebra}
[J_a,J_b]=\epsilon_{abc}J^c,\qquad
[J_a,P_b]=\epsilon_{abc}P^c,\qquad
[P_a,P_b]=\Lambda\epsilon_{abc}J^c\,,
\end{align}
admits two different non-degenerate bilinear forms. In case of EHG this would be \cite{Witten:1988hc},
\begin{align}
\tr(J_a,P_b)=\tfrac{1}{2}\eta_{ab}\,.
\end{align}
The Chern-Simons theory based on this algebra and this bilinear form can be decomposed as the sum of  two 
Chern-Simons actions of sl$(2)$ gauge algebra with opposite levels.
The conformal algebra on the other hand has a unique bilinear form. 

In this formalism, the dreibein $e^a$, and the (dualized) spin connection $\omega^a$, are 
gauge fields in the translation $P_a$ and the rotation $J_a$ generators and the 
 gauge transformations $A_\mu\to A_\mu +D_\mu\epsilon$ generate diffeomorphisms on-shell \cite{Witten:1988hc} when the gauge parameter $\epsilon$ depends linearly on fields,
 $\varepsilon^a=A^a{}_\mu\xi^\mu$,
\beqa\label{difgtn}
\delta_\xi A^a{}_\mu=\partial_\mu \xi \cdot A^a+\xi\cdot\partial A^a{}_\mu+\xi^\nu F^a{}_{\mu\nu}\,,
\eeqa

The asymptotic analysis for EHG on AdS was first done by Brown and Henneaux in \cite{Brown:1986nw} where they 
recognized that under suitable boundary conditions the asymptotic symmetries of this theory are given by two 
copies of the Virasoro algebra with the same central charge. 
A detailed analysis for CSG with AdS boundary conditions was done in \cite{Afshar:2011yh,Afshar:2011qw,Afshar:2013bla}. In the following subsection we address the main aspects of these results.

\subsection{Conformal Chern--Simons gravity}
\label{sec:2.1}
Before discussing the first order formulation of CSG as a CS gauge theory of SO(3,2), we  review the asymptotic analysis of \eqref{eq:paros3} in the metric formulation in which the metric $g$ is the dynamical field \cite{Afshar:2011yh,Afshar:2011qw}.
In Gaussian normal coordinates, consistent asymptotically locally AdS boundary conditions on the metric are,
\begin{equation}
\extd s^2= g_{\mu\nu}\extd x^{\mu} \extd x^{\nu}=e^{2\phi}\Big[\extd\rho^2+\overbrace{\left(\gamma^{(0)}_{\alpha\beta}e^{2\rho}+\gamma^{(1)}_{\alpha\beta}e^{\rho}+\gamma^{(2)}_{\alpha\beta}+\cdots\right)}^{\gamma_{\alpha\beta}}\extd x^\alpha \extd x^\beta \Big]\,,
\end{equation}
where $\rho$ is the ``radial'' coordinate and $x^\al$ the ``boundary coordinates'' (for instance, light-cone coordinates $x^\pm$).
The equations of motion \eqref{eq:paros4} impose the restrictions 
\begin{equation}\label{2nde.o.m.}
 \gamma^{(2)}_{++}=\mathcal{L}(x^+)\,,\qquad \gamma^{(2)}_{--}=\bar{\mathcal{L}}(x^-)\,\quad\text{and}\quad\,\partial_-^2\gamma^{(1)}_{++}=\gamma^{(1)}_{++}\gamma^{(2)}_{--}\,.
 \end{equation}
The most general variation of the line-element that we permit is
\eq{\begin{split}
\de \left( \extd s^2 \right)=  e^{2\phi}\left( 2 \de\phi\, \extd\rho^2  + [2\ga_{\al\be}\, \de\phi + \de \ga_{\al\be} ]\,\extd x^\al\extd x^\be \right)\, .
\end{split}
}{eq:free1}
which indicates different scenarios in deforming the boundary metric, namely the {\it trivial}, {\it fixed} and {\it free} Weyl factor, $\phi=0$, $\de\phi=0$ and $\de\phi\neq 0$, respectively. Here we consider the last case with $\phi=f(x^+,\,x^-)$ (for possible radial dependence see \cite{Afshar:2011qw}). After adding a suitable boundary term for having a well-defined variational principle,
the full on-shell variation of the action reads
\begin{equation}\label{eq:free5}
\de S_{\rm CSG}\big|_{\rm EOM} = \frac{1}{2}\,\int_{\partial M}\!\!\!\extd^2x\sqrt{-\ga^{(0)}}\,\Big(T^{\al\be}\,\de\ga^{(0)}_{\al\be} + J^{\al\be}\,\de\ga^{(1)}_{\al\be} 
\Big)\,.
\end{equation}
The response functions $T^{\al\be}$ and $J^{\al\be}$ are Brown--York stress tensor and partially massless response with conformal weights $\Delta=2$ and $\Delta=1$, respectively, whose non-zero components are given by
\begin{align}
 T_{\pm\pm} &= \mp\frac{k}{\pi}\,\left(\ga^{(2)}_{\pm\pm} +\tfrac{1}{2}\partial^2_{\pm}f\right)\label{eq:gCS48}\\
 J_{++} &= \frac{k}{2\pi}\,\ga_{++}^{(1)} \,\quad\text{with}\quad\left(\partial_-^2 -\tfrac{\pi}{k}T_{--}\right)J_{++} = 0\,. 
\end{align}
For the BTZ black hole \cite{Banados:1992wn} we obtain
\eq{
M_{\rm BTZ}= 2k r_+ r_-\,, \qquad\qquad J_{\rm BTZ}= k(r_+^2+r_-^2)\,,
}{eq:gCS59}
where $|r_+| \geq |r_-|$ are the inner and outer horizon radii, respectively (with the usual definitions of mass, $M = -\int \extd\varphi T^t_t$, and angular momentum, $J = -\int\extd\varphi T^t_\varphi$ where $x^{\pm}=t\pm \varphi$). As compared to EHG the role of mass and angular momentum is exchanged: for real $r_\pm$ the angular momentum $J_{\rm BTZ}$ is non-negative, whereas the mass $M_{\rm BTZ}$ can have either sign, exactly like in ``exotic'' gravity theories \cite{Townsend:2013ela}.

The asymptotic Weyl factor $\phi=f$ gives in general a contribution to the asymptotic charges, since CSG is only invariant under diffeomorphism and Weyl rescaling up to a boundary term. 
Conservation of the corresponding charges in turn requires cancellation of these anomalies by imposing the following conditions on the Weyl factor and its variation,
%
\eq{
\partial_+\partial_- f=0\,,\qquad\qquad \partial_t(f\partial_\varphi \de f)=\text{\small total}\;\; \varphi\text{\small -derivative}\,.
}{×}
Particularly simple choices are
 $f = f(x^+)$ 
or 
$f = f(x^-)$. 
The non-vanishing 2-point functions are given by ($z=\varphi+it$):
 \begin{align}\label{2poin}
 & \langle J_{++}(z,\bar z)J_{++}(0,0)\rangle = \frac{2k\,\bar z}{z^3}  \\
 & \langle T_{++}(z)T_{++}(0)\rangle = \frac{6k}{z^4} = -\langle T_{--}(\bar z)T_{--}(0)\rangle 
 \end{align}
These results show that one of the conformal weights of the partially massless mode is negative, $\bar h=-1/2$. This is precisely the conformal weight required for a semi-classical null state at level 2 \cite{Afshar:2011qw}, which is indeed reproduced on the gravity side through a 1-loop ghost determinant \cite{Bertin:2011jk}.
We can also read off the central charges of the dual CFT,
\eq{
c=-\bar c = 12 k \,.
}{eq:gCS54}

In order to be explicit about the derivation of the asymptotic symmetry algebra, we now move to the first order formulation where CSG can be written in terms of three  Lorentz valued variables (note that $k_{\textrm{\tiny CS}}=2k$ here), $e$, $\om$ and $\lambda$.
\begin{equation}\label{1stcnf}
S_{\textrm{\tiny CSG}}^{(1)} =\frac{k_{\textrm{\tiny CS}}}{4\pi}\int_{\mathcal{M}} \tr\left(\omega\wedge\left( d\omega 
+\tfrac{2}{3}\omega\wedge\omega\right)-2\lambda\wedge T\right) 
\end{equation}
The spin-connection is solved in terms of the dreibein $\omega=\omega(e)$ by the torsion constraint, $T=de+e\wedge\om=0$,  variation with respect to $\omega$ solves 
the Lagrange multiplier as $\lambda=S(e)$, where $S$ is the Schouten one-form, and variation with respect to $e$ gives the same field equation as in the metric formulation, $C(e)=0$ where $C$ is the Cotton one-form.
It has been shown by  Horne and Witten \cite{Horne:1988jf} that considering these variables ($e$, $\omega$ and $\lambda$) as gauge fields along translation, rotation and special conformal transformation generators and adding a 
St\"uckelberg  $\phi$ along the dilatation,
\begin{equation}\label{connection1}
A_\mu ={e^a}_\mu P_a+{\omega^a}_\mu J_a+{\lambda^a}_\mu K_a+\phi_\mu D\,,
\end{equation}
this action can be written as a Chern-Simons theory based on the SO(3,2) gauge group. 

We exploit now the Chern--Simons formulation for canonically and asymptotically analyzing CSG. 
The fact that SO(3,2) contains SO(2,2) as a subgroup, suggests that we can study 
AdS boundary conditions in this setup\footnote{The same statement holds for SO(3,1), ISO(2,1) and SO(2,1)$\times\mathbb R$ as subgroups of SO(3,2) corresponding to 
de~Sitter, Flat and Lobachevsky boundary conditions \cite{Afshar:2013bla}.}. Introducing the following state dependent one forms,
\begin{align}
 t^0&=T_1\extd t-T_2\extd\varphi\,,\qquad t^1=T_1\extd\varphi-T_2\extd t\,\qquad\text{and}\nn\\
 p^0&=P_2\extd t-P_1\extd\varphi\,,\qquad p^1=P_1\extd t-P_2\extd\varphi\,,\qquad p^2=P_3(\extd t+\extd\varphi)\,,
\end{align}
we present the AdS boundary conditions as follows \cite{Afshar:2013bla},
 \begin{equation}\label{1stb.c.s}
 \begin{split}
    e^0 &= -\ell e^{f}\left(e^{\rho}\extd t-p^0+t^0e^{-\rho}\right)\,,\qquad
    e^1 = -\ell e^{f}\left(e^{\rho}\extd\varphi-p^1-t^1e^{-\rho}\right)\,,\\
    e^2 &= -\ell e^{f}\left(\extd\rho-p^2e^{-\rho}\right)\,,\,\\
    \lambda^0 &= \tfrac{1}{2\ell}  e^{-f}\left(e^{\rho}\extd t +p^0+t^0e^{-\rho}\right)\,,\;\;\quad
    \lambda^1= \tfrac{1}{2\ell} e^{-f}\left(e^{\rho}\extd\varphi + p^1-t^1e^{-\rho}\right)\,,\\
    \lambda^2 &= \tfrac{1}{2\ell}  e^{-f}\left(\extd\rho+p^2e^{-\rho}\right)\,,\\
    \omega^0 &= e^{\rho}\,\extd\varphi + t^1e^{-\rho}\,,\qquad\qquad\qquad\;\;
    \omega^1 = e^{\rho}\,\extd t-t^0e^{-\rho}\,,\quad\quad\;\;\nn\\
    \omega^2 &= 0\,,\qquad\qquad\qquad\qquad\quad\qquad\;\;\;\;\;
     \phi=\extd f(t,\varphi)-p^2e^{-\rho}\,.
 \end{split}
 \end{equation}
Solving the flatness conditions \eqref{eq:paros2} we find, 
\begin{align}\label{{1ste.o.m.}}
 T_1=-\tfrac{1}{2}(\mathcal{L}(x^+)-\bar{\mathcal{L}}(x^-)),\qquad
 T_2=\tfrac{1}{2}(\mathcal{L}(x^+)+\bar{\mathcal{L}}(x^-)), \nonumber\\
 P_1=-P_2=\mathcal{P}(t,\varphi),\qquad P_3=\bar{\partial}\mathcal{P},\qquad \left(\bar{\mathcal{L}}-\bar{\partial}^2\right)\mathcal{P}=0\,.
\end{align}
These are the analogue of \eqref{2nde.o.m.}. A general Lie algebra-valued generator of gauge transformations is
\eq{
\varepsilon=\rho^aP_a+\tau^aJ_a+\sigma^aK_a+\gamma D.
}{BCPGTS}
The boundary conditions given in \eqref{1stb.c.s} are preserved by gauge transformations \eqref{BCPGTS} when,
  \begin{align}
 \rho^0&=\ell e^{f}\left(a_2e^{\rho}+(a_1+a_2)\mathcal{P}+a_4e^{-\rho}\right),\quad
 \sigma^0=-\tfrac{1}{2\ell} e^{-f}\left(a_2e^{\rho}-(a_1+a_2)\mathcal{P}+a_4e^{-\rho}\right)\,,\nn\\
  \rho^1&=\ell e^{f}\left(a_1e^{\rho}-(a_1+a_2)\mathcal{P}+a_3e^{-\rho}\right)\,,\quad
  \sigma^1=-\tfrac{1}{2\ell} e^{-f}\left(a_1e^{\rho}+(a_1+a_2)\mathcal{P}+a_3e^{-\rho}\right),\nn\\
\rho^2&=-\ell e^{f}\left(\partial_\varphi a_1+d_1e^{-\rho}\right)\,,\qquad\qquad\qquad
\sigma^2=\tfrac{1}{2\ell}e^{-f}\left(\partial_\varphi a_1-d_1e^{-\rho}\right)\,,\nn\\
\tau^0&=-a_1e^{\rho}+a_3e^{-\rho}\,,\qquad\tau^1=-a_2e^{\rho}+a_4e^{-\rho}\,,\qquad\tau^2=\partial_\varphi a_2\,,\qquad\nn
\gamma=\Omega+d_1e^{-\rho}\,.
   \end{align}
where the following relations should hold,
\begin{align}
 a_2&=-\tfrac{1}{2}\left(\mathcal{\epsilon}(x^+)+\bar{\mathcal{\epsilon}}(x^-)\right),\quad
 a_1=-\tfrac{1}{2}\left(\mathcal{\epsilon}(x^+)-\bar{\mathcal{\epsilon}}(x^-)\right),\quad d_1=-\bar{\partial}\mathcal{P}\epsilon(x^+)\nonumber\\
a_3&=T_2a_2-T_1a_1-\tfrac{1}{2}\partial_\varphi^2 a_1,\qquad \qquad a_4=T_1a_2-T_2a_1+\tfrac{1}{2}\partial_\varphi^2 a_2.
 \end{align}
 The variation of the state dependent functions in \eqref{1stb.c.s} with respect to these parameters are,
 \begin{align}
  \de \mathcal{L}&=\partial\mathcal{L}\epsilon+2\mathcal{L}\partial\epsilon-\tfrac{1}{2}\partial^3\epsilon\,,\qquad  
  \de \mathcal{\bar L}=\bar\partial\bar {\mathcal{L}}\bar \epsilon+2\bar {\mathcal{ L}}\bar \partial\bar \epsilon+\tfrac{1}{2}\bar {\partial}^3\bar \epsilon\,,\nn\\
   \de \mathcal{P}&=\partial\mathcal{P}\epsilon+\tfrac{3}{2}\mathcal{P}\partial\epsilon+\bar\partial\mathcal{P}\bar\epsilon-\tfrac{1}{2}\mathcal{P}\bar\partial\bar\epsilon\,,\qquad \delta_\Omega f=\Omega\,,
 \end{align}
which are the analogue of \eqref{2poin}.
The conserved charges associated to these variations are given by,
 \eq{
  Q=\frac{k_{\textrm{\tiny CS}}}{2\pi}\int \extd\varphi\, [\epsilon(x^+) \mathcal{L}(x^+)+\bar{\epsilon}(x^-)\bar{\mathcal{L}}(x^-)+\Omega(x^+)\partial_\varphi  f(x^+)]\,.
 }{}
Defining the generators of these global symmetries as,
\eq{
L_n=\tilde G[\epsilon=e^{inx^+}],\quad\quad
\bar L_n=\tilde G[\bar\epsilon=e^{inx^-}]\quad\text{and}\quad J_n=\tilde G[\Omega=e^{inx^+}]\,,
}{eq:lalapetz}
we compute the Poisson brackets and convert Poisson brackets into commutators by the prescription $i\{q,\,p\}=[\hat q,\,\hat p]$.
The resulting algebra is $\text{Vir}\oplus\overline {\text{Vir}}\oplus \hat {u}(1)_{k}$. 
Finally, we Sugawara-shift the quantum $L$ generator
\eq{
  L_m\to L_m + \frac{1}{4k} \sum_{n\in \mathbb Z} :J_n J_{m-n}:
}{eq:Lsug}

In conclusion, the asymptotic symmetry algebra has the following non-zero commutators:
\begin{align}\label{finalg1}
 [L_n,\,L_m] &= (n-m)\,L_{n+m} + \frac{c+1}{12}\,(n^3-n)\,\de_{n+m,0} \, \nn\\
 [\bar L_n,\,\bar L_m] &= (n-m)\,\bar L_{n+m} + \frac{\bar c}{12}\,(n^3-n)\,\de_{n+m,0}\, \nn\\
 [L_n,\,J_m] &= -m\,J_{n+m}\, \nn\\
 [J_n,\,J_m] &= 2k\,n\,\de_{n+m,0} \, 
\end{align}
The central charges are given by $c=-\bar c=12\,k$ with $k=k_{\textrm{\tiny CS}}/2$. 
Note the quantum shift by one in the central charge of one copy of the Virasoro algebra. This is due to the normal ordering of $J$'s introduced in \eqref{eq:Lsug}.
The relative sign of two central charges is a sign of non-unitarity. This is consistent with the parity odd nature of this theory; as mentioned before, 
flat boundary conditions seem more suitable for unitarity in the asymptotic analysis of parity odd models. 
For a detailed asymptotically flat analysis of CSG as a Chern--Simons gauge theory of SO(3,2) see \cite{Afshar:2013bla} and in the metric formulation see section \ref{sec:4} and \cite{Bagchi:2012yk}.

\subsection{Higher spin theories}
\label{sec:higher-spin-anti}

In the introduction we alluded to some similarities between CSG and higher spin theories. In this subsection we make this statement more concrete and summarize some important properties of such theories.

Even though it is easy to write down the (Fronsdal-)equations \cite{Fronsdal:1978rb} for free massless higher spin fields, the coupling of the fields for spins greater than two to gravity is severely constrained by various no-go theorems (for a review see \cite{Bekaert:2010hw}). Fradkin and Vasiliev \cite{Fradkin:1987ks} showed that consistent interacting higher spin gauge theories involving gravity need to be defined on a curved background and involve an infinite tower of massless higher spin fields \cite{Vasiliev:1992av}, see e.g.~\cite{Bekaert:2005vh,Didenko:2014dwa} for reviews.

One interesting aspect of higher spin gauge fields is that they might be connected to string theory in the tensionless limit in which the massive excitations of string theory become massless. It is conjectured that string theory is a broken phase of a higher spin gauge theory. For more details see \cite{Sagnotti:2011qp} and references therein.

Another interesting aspect is that holographic correspondences between higher spin theories and field theories can be formulated, such as the conjectured duality in the large $N$ limit of the critical 3-dimensional $O(N)$ model and the  minimal bosonic higher spin theory in $\mathrm{AdS}_{4}$ \cite{Klebanov:2002ja,Sezgin:2002rt,Sezgin:2003pt} (for a review of various impressive checks of this conjecture see \cite{Giombi:2009wh}).

We focus now on $2+1$ dimensions where the situation simplifies significantly. 
An action is known \cite{Blencowe:1988gj}, namely the sum of two Chern--Simons actions \eqref{eq:paros1} with opposite levels with the gauge group $SL(N)$ which is a natural generalization of EHG and corresponds to fields of spin $s=3,4,\ldots,N$ coupled to gravity. 
This consistent truncation to a finite number of higher spin fields is not possible in higher dimensions \cite{Aragone:1979hx}. Moreover, the dual field theories are 2-dimensional, which allows a high degree of analytic control.

The Brown--Henneaux type of analysis reviewed in the previous subsection generalizes to higher spin fields for asymptotic $\mathrm{AdS_{3}}$ \cite{Henneaux:2010xg,Campoleoni:2010zq,Gaberdiel:2011wb,Campoleoni:2011hg} and leads to asymptotic $\mathcal{W}_{N} \times \mathcal{W}_{N}$ \cite{Zamolodchikov:1985wn,Bouwknegt:1992wg} symmetry algebras.
Using the infinite dimensional higher spin algebras $hs[\lambda] \oplus hs[\lambda]$ as gauge algebra we get gravity coupled to massless fields with spins $s=3,4,\ldots,\infty$ and, again for $\mathrm{AdS_{3}}$, asymptotic symmetries of the form  $\mathcal{W}_{\infty}[\lambda] \times \mathcal{W}_{\infty}[\lambda]$.

Gaberdiel and Gopakumar proposed \cite{Gaberdiel:2010pz} that the $hs[\lambda]$ theory coupled to additional massive scalar fields on $\mathrm{AdS}_{3}$ is dual to a specific large-$N$ limit of $\mathcal{W}_{N}$ minimal models on the $\mathrm{CFT}$ side. The duality is reviewed in \cite{Gaberdiel:2012uj}.

Since the BTZ black hole can also be generalized to higher spin theories, new questions arise concerning gauge invariant characterizations of observables --- like in CSG there are gauge symmetries that act on the metric but are not diffeomorphisms --- and black hole thermodynamics (for a review of the proposed answers see \cite{Ammon:2012wc,Perez:2014pya}).

An interesting possibility that we will exhibit in the next section --- first for CSG and then for higher spin theories --- is to realize higher spin holography for backgrounds other than AdS$_{3}$~\cite{Gary:2012ms}, see \cite{Afshar:2012nk,Riegler:2012fa,Afshar:2013vka,Gonzalez:2013oaa} for explicit constructions.

\section{Lobachevsky holography}
\label{sec:3}

Lobachevsky holography refers to asymptotic expansions of the line-element of the form
\eq{
\extd s^2=\pm \extd t^2 + \extd\rho^2 +\sinh^2\rho\extd\varphi^2 + \dots
}{eq:paros5}
where the ellipsis refers to suitable expressions subleading as $\rho\to\infty$. 
Without subleading terms the line-element \eqref{eq:paros5} describes a direct product manifold of the 2-dimensional Lobachevsky plane $\mathbb{H}_2$ (famously depicted by M.C.~Escher in his paintings ``Circle Limits'') and a line or $S^1$ corresponding to the time-direction (with upper sign: Euclidean time).
Which subleading expressions are ``suitable'' depends on the specific theory.

In \cite{Bertin:2012qw} boundary conditions suitable for CSG were formulated and their consistency was checked. Performing the Brown--Henneaux type of analysis reviewed in section \ref{sec:2.1} then leads to the asymptotic symmetry algebra 
\begin{align}
\label{eq:paros6}
 [L_n,\,L_m] &= (n-m)\,L_{n+m} + \frac{c}{12}\,(n^3-n)\,\de_{n+m,0} \nn\\
 [L_n,\,J_m] &= -m\,J_{n+m}\nn\\
 [J_n,\,J_m] &= 2k\,n\,\de_{n+m,0} \, .
\end{align}
The value of the central charge, $c=24k$, is compatible with the limiting case of warped AdS holography \cite{Anninos:2008fx}.
The algebra above is similar to the AdS asymptotic symmetry algebra \eqref{finalg1}, with the following differences: there is no second copy of the Virasoro algebra and no quantum shift by one in the central charge. The appearance of a single Virasoro algebra and a $\hat{u}(1)$ current algebra suggests that the dual field theory, if it exists, is a warped CFT \cite{Detournay:2012pc}. Some checks and aspects of this proposal --- consistency of canonical charges, one-loop partition function, identification of non-perturbative states, aspects of the Lobachevsky $\leftrightarrow$ field theory map --- are discussed in \cite{Bertin:2012qw}, but many open issues remain (some of which are also mentioned in that paper).

Amusingly, the higher spin side of the Lobachevsky story seems more straightforward, so let us switch now to higher spin theories. The first explicit example of non-AdS holography was worked out in \cite{Afshar:2012nk} for spin-3 gravity (for more details see \cite{Riegler:2012fa}). In this example one considers a bulk metric that is asymptotically $\mathbb{H}_2\times\mathbb{R}$. 
In order to succeed it is crucial that the embedding of $\mathfrak{sl}(2)$ into  $\mathfrak{sl}(3)$  yields at least one singlet under the $\mathfrak{sl}(2)$. Otherwise it turns out that one cannot reproduce the correct $\extd t^2$ term in the line-element \eqref{eq:paros5}. The unique viable choice for spin-3 gravity is then the non-principal embedding of $\mathfrak{sl}(2)$ into  $\mathfrak{sl}(3)$ (also called diagonal embedding). In this way we reproduce \eqref{eq:paros5} (up to subleading terms) in the limit $\rho\rightarrow\infty$. 

Besides the $\mathfrak{sl}(2)$ part given by the generators $L_i$ with $i=0,\pm1$ this embedding contains the singlet $S$ and ``colored'' doublets $\psi^\pm_{j}$ with $j=\pm\frac{1}{2}$. We write the connections as
	\begin{equation}
		a_\mu=\hat{a}_\mu^{(0)}+a_\mu^{(0)}+a_\mu^{(1)}\quad\textnormal{and}\quad\bar{a}_\mu=\hat{\bar{a}}_\mu^{(0)}+\bar{a}_\mu^{(0)}+\bar{a}_\mu^{(1)}\,.
	\end{equation}
One set of connections reproducing \eqref{eq:paros5} in the large $\rho$ limit is given by
	\begin{subequations}\label{eq:H2xRBCs}
		\begin{align}
			\hat{a}_\rho^{(0)}=&L_0,\quad\hat{a}_\varphi^{(0)}=-\frac{1}{4}L_1,\quad\hat{\bar{a}}_\rho^{(0)}=-L_0,\quad\hat{\bar{a}}_\varphi^{(0)}=-L_{-1},\quad\hat{\bar{a}}_t^{(0)}=\sqrt{3}S\\
			a_\varphi^{(0)}=&\frac{2\pi}{k}\left(\frac{3}{2}\mathcal{W}_0(\varphi)S+\mathcal{W}^+_{\frac{1}{2}}(\varphi)\psi^+_{-\frac{1}{2}}-\mathcal{W}^-_{\frac{1}{2}}(\varphi)\psi^-_{-\frac{1}{2}}-\mathcal{L}(\varphi)L_{-1}\right),\\
			\bar{a}_\varphi^{(0)}=&\frac{2\pi}{k}\left(\frac{3}{2}\bar{\mathcal{W}}_0(\varphi)S+\bar{\mathcal{W}}^+_{\frac{1}{2}}(\varphi)\psi^+_{-\frac{1}{2}}+\bar{\mathcal{W}}^-_{\frac{1}{2}}(\varphi)\psi^-_{-\frac{1}{2}}+\bar{\mathcal{L}}(\varphi)L_{-1}\right),\\
			\hat{a}_t^{(0)}=&a_\rho^{(0)}=a_t^{(0)}=\bar{a}_\rho^{(0)}=\bar{a}_t^{(0)}=0,\\
			a^{(1)}_\mu=&\mathcal{O}(e^{-2\rho})=\bar{a}^{(1)}_\mu,
		\end{align}
	\end{subequations}
where the $\hat{a}_\mu^{(0)}$ $(\hat{\bar{a}}_\mu^{(0)})$ describe the part of the connection that reproduces the background, $a_\mu^{(0)}$ $(\bar{a}_\mu^{(0)})$ state dependent fluctuations that are of leading order for large $\rho$ and $a_\mu^{(1)}$ $(\bar{a}_\mu^{(1)})$ are subleading terms. 

As in the example in section \ref{sec:2.1}, in order to check whether or not the boundary conditions lead to interesting physics one has to find gauge transformations that preserve these boundary conditions and check that the resulting canonical boundary charge is finite at the boundary, nontrivial and conserved in time. After having determined a canonical boundary charge which satisfies these conditions one can determine the asymptotic symmetry algebra on the level of Poisson brackets. One can then replace $i\{\cdot,\cdot\}\rightarrow[\cdot,\cdot]$ and expand the fields appearing in \eqref{eq:H2xRBCs} in terms of their Fourier modes in order to obtain the (semi-classical) symmetry algebra which determines essential properties of the dual quantum field theory. 

In the case of the boundary conditions \eqref{eq:H2xRBCs} the asymptotic symmetry algebra obtained this way consists of one copy of the semi-classical (large values of $k_{\textrm{\tiny CS}}$) $\mathcal{W}^{(2)}_3$ algebra, also known as Polyakov-Bershadsky Algebra \cite{Bershadsky:1990bg,Polyakov:1989dm} and one copy of an affine $\hat{\mathfrak{u}}(1)$ algebra. This is the anticipated spin-3 generalization of the CSG result \eqref{eq:paros6}.

Since the $\mathcal{W}^{(2)}_3$ algebra is an infinite dimensional, non-linear, centrally extended algebra one has to introduce normal ordering prescription for the non-linear terms if we are interested in the regime where $k_{\textrm{\tiny CS}}$ is of order one, i.e., in the quantum regime. The structure constants of the $\mathcal{W}^{(2)}_3$ algebra are functions of $k_{\textrm{\tiny CS}}$. Hence one has to check whether or not the algebra still satisfies the Jacobi identities after introducing normal ordering. And indeed, in order to be compatible with the Jacobi identities, some of the structure constants and the central charges obtain $\mathcal{O}(1)$ corrections in the quantum regime. The final result for the asymptotic symmetry algebra for connections obeying \eqref{eq:H2xRBCs} is $\mathcal{W}^{(2)}_3\oplus\hat{\mathfrak{u}}(1)$.

After having found the quantum asymptotic symmetry algebra of spacetimes that are asymptotically $\mathbb{H}_2\times\mathbb{R}$ one can also ask whether or not there are unitary representations of this algebra. In the case of Lobachevsky holography it is surprisingly easy to answer this question. There is only one value of the Chern Simons level $k_{\textrm{\tiny CS}}$ where it is possible to obtain nontrivial unitary representations \cite{Afshar:2012nk,Riegler:2012fa}. The reason why this question is so easy to answer in this case is because the states that correspond to descendants of the ``colored'' doublet have to be absent, otherwise those states would always have norms with opposite signs spoiling unitarity. This leaves only two possible values of the level $k_{\textrm{\tiny CS}}$ with only one of them leading to a nontrivial theory, which can be interpreted as the theory of a free boson with a coupling constant fixed by an additional gauge symmetry. The generalization of the unitarity discussion to the full $\mathcal{W}^{(2)}_N$ family is more involved, particularly for even $N$ \cite{Afshar:2014cma}.

\section{Flat space holography}
\label{sec:4}

The constructions reviewed above are all similar at a technical level. This has two reasons. First, we were always dealing with some Chern--Simons theory \eqref{eq:paros1} supplemented by suitable boundary conditions (finding the latter was the main non-trivial task). Second, we were almost exclusively concerned with asymptotic symmetry algebras and did not specify in detail the precise field theory that is supposed to be dual to a given gravitational or higher spin theory, other than that it has to fall into representations of the corresponding asymptotic symmetry algebra (given that all these symmetry algebras are infinite dimensional and have specific values of the central charges predicted from the gravity calculation this puts already a lot of constraints on the dual 2-dimensional field theory). In addition, all the constructions above referred to some curved asymptotic background.

In this section we go beyond this basic scenario, by allowing for non-topological theories like topologically massive gravity, by attempting to establish a more precise holographic correspondence to specific field theories, and by studying backgrounds that are locally and asymptotically flat. In section \ref{sec:4.1} we review attempts to establish precise holographic correspondences between AdS quantum gravity and specific CFTs, before addressing the flat case in section \ref{sec:4.2}, where we shall come back to our starting point, CSG.

\subsection{Introduction to 3-dimensional quantum gravity in AdS}\label{sec:4.1}

Quantum gravity is a notoriously difficult subject. As such, one strategy to tackle it is to consider toy models capturing some of its salient features. EHG in (2+1)-dimensions has emerged over the years as an archetypical model for quantum gravity in general, and AdS/CFT in particular. It differs in important respects from its (3+1)-dimensional counterpart: it has no bulk propagating degrees of freedom, and any solution to the equations of motion has constant curvature (i.e. is flat for vanishing cosmological constant $\Lambda := -1/\ell^2$; for reviews, see e.g.~\cite{Carlip3,Carlip4,Carlip:1998uc}, and \cite{0410294} p.~29 for a chronological list of references). Despite the remarkable observation that 3-dimensional gravity could itself be formulated as a Chern-Simons theory of the form (\ref{eq:paros1})  \cite{Achucarro:1987vz, Witten:1988hc,Witten:1989sx} with a gauge group depending on $\Lambda$, it appeared at first sight too simple to be able to address the conundrums of quantum gravity.
The situation changed dramatically through a series of seminal contributions in the negatively curved case $\Lambda < 0$ of which we cite three hereafter. 

First, even though there are no \emph{bulk} degrees of freedom, the presence of an asymptotic boundary in AdS$_3$ induces \emph{boundary} degrees of freedom \cite{Carlip4}. In particular, the phase space of AdS$_3$ gravity admits a non-trivial action of the 2-dimensional conformal group with two sets of non-trivial Virasoro charges $L_n^\pm$ and non-vanishing central charge given by $c^\pm = \frac{3 \ell}{2 G}$. This appeared as the first hint of a deep connection between a gravity theory in AdS space and a conformal field theory in one dimension less. 

Second, the AdS$_3$ phase space happens to contain black hole solutions, the BTZ black holes \cite{BTZ, BHTZ} with the exciting prospect of addressing questions related to black hole physics in a simplified setting. 

Third, assuming the existence of a dual CFT$_2$ of which BTZ black holes are particular thermal 
states, the BTZ Bekenstein-Hawking entropy could be reproduced by a counting of states using the Cardy formula\cite{Strominger}. 

Despite these striking and suggestive results, the precise nature of the corresponding dual CFT$_2$ (in pure gravity) remained elusive for another 10 years. In 2007, Witten revisited the subject and made a concrete proposal for the partition function of the CFT dual to pure 3-dimensional gravity \cite{Witten:2007kt}. Assuming holomorphic factorization (motivated partially by the relation to Chern-Simons theory), he argued from the BTZ spectrum in AdS$_3$ gravity that the holomorphic part of the partition function should take the form (with $k = c/24$ quantized to integers)
\beq \label{ExtremalCFT}
   Z(q) = \sum_{r=0}^{k} a_r J(q)^r, \quad J(q) = \frac{1}{q} + 196884 q + \cdots
\eeq
where $J(q)$ is the unique modular-invariant function on the upper half plane, which is holomorphic away from a single pole at the cusp. Therefore, the requirement that the partition function be of the form 
\beq
  Z(q)  = Z_0 (q) + O(q), \quad Z_0 (q) = q^{-k} \prod_{n=2}^{\infty} \frac{1}{1 - q^n}\,,
\eeq
where $Z_0 (q)$ captures the vacuum descendants and the ``$O(q)$'' piece the BTZ black holes (having $L_0 > 0$), uniquely fixes the form of the partition function. CFTs with partition functions \re{ExtremalCFT} are called \emph{extremal}, roughly because they have as few low-lying primaries as possible compatible with modular invariance, and display remarkable group- and number-theoretic properties.

It happens that AdS$_3$ gravity is simple enough that the quantum gravity partition function can be explicitly calculated as a sum over geometries. Maloney and Witten performed this computation \cite{Maloney-Witten} and found out  that the result could \emph{not} be interpreted as a CFT partition function, i.e., as a trace over some CFT Hilbert space. They concluded that either pure gravity in 2+1 dimensions simply did not exist quantum mechanically, or that additional contributions should be included. At any rate, the quantity they computed did not holomorphically factorize, thereby violating one of the assumptions of \cite{Witten:2007kt}. 

An alternative emerged few months later under the name \emph{chiral gravity} \cite{LSS,MSS}. The idea was to modify pure gravity by supplementing if with the gravitational Chern--Simons term \re{eq:paros3}. The resulting theory is called \emph{Topologically Massive Gravity} (TMG) \cite{Deser:1982vy, Deser:1981wh} with action
\beq \label{tmgaction}
S_{\textrm{\tiny TMG}} = \frac{1}{16 \pi}\,\int \extd^3 x \sqrt{-g}\,\big(R + \frac{2}{\ell^2}\big) - \frac{1}{8k \mu}\,S_{\textrm{\tiny CSG}}\,.
\eeq
One effect of the additional term \eqref{eq:paros3} is to shift the values of the (asymptotically) conserved charges as compared to EHG. For Brown--Henneaux boundary conditions \cite{TMGHenneaux0901}
\beqa
\Delta g_{rr} = \frac{f_{rr}}{r^4} + {\cal O}(\frac{1}{r^5}) \qquad \Delta g_{r\pm} = \frac{f_{r\pm}}{r^3} + {\cal O}(\frac{1}{r^4}) \qquad \Delta g_{\pm\pm} = f_{\pm\pm} + {\cal O}(\frac{1}{r})
\eeqa
the corresponding Virasoro charges are given by
\beq\label{VirCharges}
  L_n^{\pm} = \frac{2}{\ell} \left(1 \pm \frac{1}{\mu \ell}\right) \int \, e^{i n x^{\pm}} f_{\pm \pm} d\phi
\eeq
with the corresponding central extensions \cite{Kraus:2005zm}
\beq
   c^{\pm} =  \left(1 \pm \frac{1}{\mu \ell}\right) \frac{3 \ell}{2 G}\,.
\label{eq:critical}
\eeq
Therefore, at the critical point $\mu \ell = 1$, one copy of the Virasoro algebra has vanishing central charge. If the theory is unitary then it must be chiral and one is left with a single copy of the Virasoro algebra. Alternatively, if the theory is non-unitary one encounters the structure of a specific type of logarithmic CFT where one chiral part of the stress tensor acquires a logarithmic partner \cite{Grumiller:2008qz,Grumiller:2013at}. In the former case, holomorphic factorization would be explicitly implemented in the resulting theory, dubbed ``chiral gravity'' \cite{Li:2008dq} (see also \cite{Maloney:2009ck}). Chiral gravity (which could exist as a unitary truncation of the non-unitary logarithmic CFT that is dual to TMG at the critical point  $\mu \ell = 1$) therefore appears as a candidate for the simplest and potentially solvable model including quantum black holes.

\subsection{Flat space chiral gravity}\label{sec:4.2}

The above considerations regarded gravity theories with a negative cosmological constant. Could a similar logic be used to argue that flat space could be dual to a field theory of some kind? And if yes, what could it be?

It is tempting to use as guiding principle the ingredients that led to the first glimpses of AdS/CFT: asymptotic symmetries. The first caveat is that the asymptotic structure of flat space is more involved than that of AdS spaces (see e.g.~\cite{waldgeneral}). However, the structure of its various asymptotic symmetry groups has been studied over the years, starting with \cite{Bondi:1962}. For the case that will interest us in the following, the asymptotic symmetries of (2+1)-dimensional gravity at null infinity form the so-called BMS$_3$ algebra \cite{Barnich:2006av}, with commutation relations
\begin{subequations}
\label{eq:GCA} 
\begin{align}
[L_m,\,L_n]&= (m-n)\,L_{n+m} + \frac{c_1}{12}\,(n^3-n)\,\de_{n+m,0}  \label{eq:virasoro} \\
[L_m,\,M_n]&= (m-n)\,M_{n+m} + \frac{c_2}{12}\,(n^3-n)\,\de_{n+m,0} \\
[M_m,\,M_n]&= 0
\end{align}
\end{subequations}
It is generated by Virasoro generators $L_n$ and supertranslations $M_n$. The latter are the modes of diffeomorphisms preserving the following boundary conditions at null infinity \cite{Bagchi:2012yk} (the functions $h_i$ depend on $\theta$; the functions $h_{\mu\nu}$ depend on $u$ and $\theta$):
\begin{subequations}
 \label{BMSBC}
\begin{align}
 g_{uu} &= h_{uu} + O(\tfrac 1r) \qquad g_{ur} = -1 + h_{ur}/r + O(\tfrac{1}{r^2}) \\ 
 g_{u\theta} &= h_{u\theta} + O(\tfrac 1r) \qquad g_{rr} = h_{rr}/r^2+O(\tfrac{1}{r^3}) \\ 
 g_{r\theta} &= h_1   + h_{r\theta}/r + O(\tfrac{1}{r^2}) \\
 g_{\theta\theta} &= r^2 + (h_2 + u h_3) r + O(1)
\end{align}
\end{subequations}
The flat counterpart of \re{VirCharges} is then given by 
\begin{subequations}
\label{ChargesBMSTMG}
 \begin{align}
  M_n &= \frac{1}{16 \pi G}\, \int \extd\theta\, e^{i n \theta}
 \; \big(h_{uu} + h_3\big)   \\
  L_n &= \frac{1}{16 \pi G \mu}\, \int\extd\theta\, e^{i n \theta} \; \big(h_{uu}  + h_3\big)
\nonumber\\
& \!\!\!\!+ \frac{1}{16 \pi G}\, \int \extd\theta\, e^{i n \theta}
 \; \big( i n u h_{uu} + i n h_{ur} + 2 h_{u\theta} + \partial_u h_{r\theta} 
- h_3  h_1 - i n \partial_\theta h_1      \big)          
\end{align}
\end{subequations}
and the central extensions in \re{eq:GCA} are computed as\cite{Bagchi:2012yk}\footnote{%
It should be straightforward to generalize these results to other massive gravity theories like ``new'' massive gravity \cite{Bergshoeff:2009hq,Bergshoeff:2009aq}. 
}
\beq\label{BMSCentral}
c_1 = \frac{3}{\mu G}\,, \qquad c_2 = \frac{3}{G}\,.
\eeq

The phase space defined by the boundary conditions \re{BMSBC} contains an interesting two-parameter family of solutions recognized some time ago as the shift-boost orbifold of flat space \cite{Cornalba:2002fi}:
\begin{equation}\label{SB}
 \extd s^2 = 8 m \, \extd u^2  -2\extd r\extd u + 8 j \, \extd\theta \extd u + r^2 \extd\theta^2 .
\end{equation}
They represent cosmological solutions (here expressed in Eddington--Finkelstein coordinates) --- in particular, they have a cosmological horizon, an associated Bekenstein--Hawking entropy and a Hawking temperature \cite{Bagchi:2012xr, Barnich:2012xq}. We therefore have a classical phase space endowed with an action of an infinite-dimensional BMS$_3$ symmetry, and by analogy with the AdS$_3$ situation, one could expect that upon quantization states will form representation of that algebra, i.e. quantum gravity in flat space would be related to a BMS$_3$-invariant field theory. Although some hints in this direction have been given, it is fair to say these types of field theories remain relatively unexplored. Some aspects of the representation theory have been discussed in \cite{Bagchi:2009my, Bagchi:2010zz, 1210.0731, 1403.3835, 1403.5803}. What is lacking as opposed to the exhaustive study of 2-dimensional CFTs is the presence of concrete examples of such field theories. We review now a first concrete example of holography in flat spacetimes. 

To this end, there is a limit that make our lives easier. Consider 
\beq
\mu \to 0\,, \quad G \to\infty \quad \mbox{keeping} \quad \mu G := \frac{1}{8k} \quad \mbox{finite}. 
\eeq 
In that limit, the $M_n$ charges become trivial, the central term $c_2$ vanishes and the BMS$_3$ algebra reduces to a single copy of a Virasoro algebra! This can be further checked by looking at null vectors in the field theory and observing that in the above limit, there is indeed a consistent truncation of the representations of the algebra \re{eq:GCA} to simply the Virasoro module \cite{Bagchi:2012yk}. 
On the bulk side, the Bekenstein-Hawking entropy of the above solutions (taking into account the Chern-Simons contribution \cite{Solodukhin:2005ah, Kraus:2005zm, Bouchareb:2007yx, Tachikawa:2006sz}) is 
\beq
   S = 8 \pi k \sqrt{2 m} = 2 \pi \sqrt{\frac{c_1 L_0}{6}}
\eeq
i.e., precisely a chiral half of the Cardy formula. This provides a check on the correctness of flat space holography. 

One can go further. The vacuum flat space solution lies in \re{SB} for $m=-\frac{1}{8}$ and $j=0$, i.e., for $L_0 = -k = -\frac{c}{24}$, while the cosmological solutions have $L_0 > 0$. The spectrum therefore share strong similarities with that of AdS$_3$ gravity, as there is a gap between the vacuum and the first primary state. One can then follow the same reasoning as Witten, arguing that modular invariance uniquely fixes the partition function to be of the form \re{ExtremalCFT}. As a consequence, we can proceed with a comparison analogous to the one done in \cite{Witten:2007kt} for BTZ black holes. Consider a cosmological solution with $L_0 = 1$, at $k=1$. Its (semi-classical) entropy is $S_{\textrm{BH}} = 4 \pi \sim 12.57$. On the other hand, in the expansion \re{ExtremalCFT}, 196884 is the total number of states with $L_0 = 1$, representing one descendant of the vacuum state and 198883 primaries creating the corresponding cosmological solution. The entropy is thus $S_{\textrm{\tiny CFT}} = \ln 196883 \sim 12.19$, which matches with the geometrical entropy within a few percents (perfect agreement was not be expected since the semi-classical entropy is valid for large $k$ and we used $k=1$; the agreement gets better as $k$ increases). This leads us to conjecture that CSG with the above boundary conditions --- a theory which we call flat space chiral gravity --- is dual to a chiral CFT with $c = 24 k$. 

This conjecture can be sharpened by further arguments, which we now present.
The presence of the finite sized gap leads to the expectation that the dual CFT is an extremal CFT with $c=24k$.   An important caveat is that such CFTs need not exist for arbitrary values of $k$ \cite{Gaberdiel:2007ve,Gaberdiel:2008pr}, but at least for $k=1$ the extremal CFT that could serve as a gravity dual has been previously identified by Witten \cite{Witten:2007kt} as the Monster CFT \cite{Frenkel:1985vk}. So we can sharpen our conjecture to the following \cite{Bagchi:2012yk}: 

\vspace{0.1cm}

\noindent{\em {Flat space chiral gravity at Chern--Simons level $k=1$ is dual to the Monster CFT.}}







\begin{acknowledgement} 
HA was supported by the Dutch stichting voor Fundamenteel Onderzoek der Materie (FOM). AB was supported by an INSPIRE award of the Department of Science and Technology, India. 
SD is a Research Associate of the Fonds de la Recherche Scientique F.R.S.-FNRS (Belgium). 
DG and SP were supported by the START project Y~435-N16 of the Austrian Science Fund (FWF) and the FWF projects I~952-N16 and I~1030-N27. MR was supported by the Austrian Science Fund (FWF) and the FWF project I~1030-N27.

DG dedicates this proceedings contribution to the memory of his grandmother Gerda.
\end{acknowledgement}

\newpage


\providecommand{\href}[2]{#2}\begingroup\raggedright\endgroup

\end{document}